\documentclass[aps,prl,showpacs,twocolumn,
nofootinbib]{revtex4}
\usepackage[dvips]{epsfig}
\usepackage{graphicx}
\usepackage{amsmath}

\begin{document}

\draft


\title{ \hfill\vbox{\hbox{\small UMD-PP-05-024 \qquad\qquad}}\\ \Large\bf
Leptogenesis and $\mu-\tau$ symmetry }
\author{\bf  R.N. Mohapatra  and S. Nasri }

\affiliation{ Department of Physics, University of Maryland, College Park,
MD 20742, USA}

\date{October, 2004}

\begin{abstract}
If an exact $\mu\leftrightarrow \tau$ symmetry is the explanation of
the maximal atmospheric neutrino mixing angle, it has interesting
implications for the origin of matter via leptogenesis in models where
small neutrino masses arise via the seesaw mechanism. For
seesaw models with two right handed neutrinos $(N_\mu, N_\tau)$, lepton
asymmetry vanishes in the exact  $\mu\leftrightarrow \tau$  symmetric
limit, even though there are nonvanishing Majorana phases in the neutrino
mixing matrix. On the other hand, for three right handed neutrino models,
 lepton asymmetry is nonzero and is given directly by the solar mass
difference square. We also find an upper bound on the lightest neutrino
mass.
\end{abstract}


\maketitle

\section{Introduction}
One of the most puzzling aspects of neutrino mixings observed in
various oscillation experiments is the near maximal value of the
$\nu_\mu-\nu_\tau$ mixing angle (i.e. $\theta_{23}\simeq \pi/4$). This was
needed to explain the original atmospheric neutrino data and is now
supported by data from the K2K experiment that uses accelerator
neutrinos. The corresponding parameter
in the quark sector is very small (about $4\%$) and is believed to be
connected to the mass hierarchy among quarks. The large value of
$\theta_{23}$ may therefore be telling us about  some new symmetries
of leptons that are not present in the quark sector and may provide
a clue to understanding the nature of quark-lepton
physics beyond the standard model.

To explore this further, the first step is to write down the
neutrino mass matrix that leads to a near maximal $\theta_{23}$ and
then try to see what physics leads to it. It is well
known that the neutrino mixings are a combined effect of the structure of
both the charged lepton and the neutrino mass matrices. If we write
\begin{eqnarray}
{\cal L}_m~=~\nu^T_\alpha C^{-1}{\cal M}_{\nu,\alpha\beta}\nu +
\bar{e}_{\alpha,L}M^e_{\alpha \beta}e_R + h.c.,
\end{eqnarray}
diagonalizing the mass matrices by the transformations
$U^T_\nu{\cal M}_\nu U_\nu={\cal M}^\nu_{diag}$ and $U^{\dagger}_\ell
M^eV~=~M^e_{diag}$, gives the lepton mixing matrix
$U_{PMNS}~=~U^{\dagger}_\ell U_\nu$.
It is conventional to parameterize $U_{PMNS}$ in terms of three angles
$\theta_{12}$
(the solar angle), $\theta_{23}$ (the atmospheric angle) and $\theta_{13}$
the reactor angle as well as three phases. Our goal is to understand the
near maximal value of $\theta_{23}$ using a leptonic symmetry and study
its implications.

 A fundamental theory
can of course determine the structure of both the charged lepton and the
neutrino mass matrices and
therefore will lead to predictions about lepton mixings. However, in the
absence of
such a theory, if one wants to adopt a model independent approach and look
for symmetries
that may explain the value of $\theta_{23}$, it is useful to work in a basis
where charged leptons are mass eigenstates
 and hope that any symmetries for leptons revealed in this
basis are true or approximate symmetries of Nature.

 In the basis where charged leptons are mass eigenstates, a symmetry that
has proved useful in understanding maximal atmospheric neutrino mixing is
 $\mu\leftrightarrow \tau$ interchange symmetry\cite{mutau}. The mass
difference between the muon and the tau lepton of course breaks this
symmetry. So we expect this symmetry to be an approximate one. It may
however happen that the symmetry is truly exact at a very high scale; but
at low mass scales, the effective theory only has the $\mu-\tau$ symmetry
in the
neutrino couplings but not in the charged lepton sector so that we
have $m_\tau \gg m_\mu$\cite{grimus}. We will consider
this class of theories in this note.
For this case, a convenient parameterization of the neutrino mass
matrix
is (assuming the neutrinos to be Majorana fermions):
\begin{eqnarray}
{\cal M}_\nu~=~\frac{\sqrt{\Delta
m^2_A}}{2}\left(\begin{array}{ccc}c\epsilon^n
&d\epsilon &d\epsilon\\ d\epsilon & 1+\epsilon & -1 \\
d\epsilon & -1 & 1+\epsilon\end{array}\right)
\end{eqnarray}
where $n\geq 1$.
An immediate prediction of this mass matrix is that $\theta_{23}=\pi/4$
and $\theta_{13}=0$; we also get $\epsilon~\sim \sqrt{\Delta
m^2_\odot/\Delta m^2_A}$.

 We can now use
$\theta_{13}$ as a probe of how leptonic $\mu\leftrightarrow
\tau$ symmetry is broken in Nature and through that one may hope for an
understanding of the origin of the near maximal (maximal ?) $\theta_{23}$,
as has been emphasized in ref.\cite{recent} (and also perhaps the
$\mu-\tau$ mass difference). In
particular, different
ways of breaking  $\mu\leftrightarrow \tau$ symmetry will lead to
$\theta_{13}\sim \sqrt{\Delta m^2_\odot/\Delta m^2_A}$ or
$\theta_{13}\sim {\Delta m^2_\odot/\Delta m^2_A}$. These predictions are
clearly  timely and interesting in view of many proposals to measure the
parameter $\theta_{13}$\cite{reactor,bnl}\footnote{The $\mu-\tau$
symmetry in supersymmetric seesaw models also leads to other
phenomenological predictions such as
the $B(\mu\rightarrow e+\gamma)/B(\mu\rightarrow e \nu \bar{\nu})=
B(\tau\rightarrow e+\gamma)/B(\tau\rightarrow e \nu \bar{\nu})$ .}.

In this paper, we discuss implications of exact $\mu\rightarrow \tau$
symmetry
for the origin of matter via leptogenesis\cite{yana} and find several
new results:
(i) we find that if there are only two right handed neutrinos
$(N_\mu,N_\tau)$ that via seesaw mechanism lead to neutrino masses,
then primordial lepton asymmetry arising from right handed
neutrino decay vanishes in the $\mu-\tau$ symmetric limit even though in
the low energy neutrino mass matrix
may have Majorana phases; (ii) secondly, for the case of three right
handed neutrinos,  the primordial lepton asymmetry is
directly proportional to the solar mass difference square. These
predictions are very different from the generic three neutrino
case\cite{buch}. In both these case we assume
that neutrino masses arise via the type I seesaw formula. These results
are independent of any detailed model.

\section{Primordial lepton asymmetry with two righthanded neutrinos}
We start with the neutrino part of the superpotential:
\begin{eqnarray}
W~=~e^{cT}{\bf Y_{\ell}}L H_d+ {N^c}^T{\bf Y_{\nu}}L H_u+\frac{1}{2} {\bf
M_R}{N^c}^TN^c
\end{eqnarray}
where we assume that $N^c\equiv (N^c_\mu,N^c_\tau)$.
As noted earlier, we work in a basis where ${\bf Y_\ell}$ is
diagonal. While naively, one may think that in such models
$m_\mu=m_\tau$, there are
models where one can split the muon and tau masses
consistent with this symmetry in the neutrino sector\cite{grimus}.

The basic assumption of this work is that we have models where
${\bf Y_\nu}$ and ${\bf M_R}$ obey $\mu\leftrightarrow \tau$ symmetry
under which
$(N_\mu\leftrightarrow N_\tau)$ and $L_\mu\leftrightarrow L_\tau$
whereas the $m_\mu\neq m_\tau$. The
general structure of ${\bf Y_\nu}$ and ${\bf M_R}$ are then given by:
\begin{eqnarray}
{\bf M_R}~=~\left(\begin{array}{cc}M_{22} & M_{23}\\ M_{23} &
M_{22}\end{array}\right)\\ \nonumber
{\bf Y_\nu}~=~\left(\begin{array}{ccc}h_{11} & h_{22} &  h_{23} \\
h_{11} & h_{23} & h_{22}\end{array}\right)
\end{eqnarray}
The seesaw formula in our notation is
\begin{eqnarray}
{\cal M}_{\nu}~=~-{\bf Y^T_\nu M^{-1}_RY_{\nu}}
{v^2_{wk}}
\label{seesaw}\end{eqnarray}
and the formula for primordial lepton asymmetry in this case, caused by
right handed
neutrino decay is\cite{lepto}
\begin{eqnarray}
\epsilon_1~=~\frac{1}{4\pi}\sum_j\frac{ Im [\tilde{Y}_\nu
\tilde{Y}^{\dagger}_\nu]^2_{12}}{(\tilde{Y}_\nu
\tilde{Y}^{\dagger}_\nu)_{11}}
F(\frac{M_1}{M_2})
\label{nl}\end{eqnarray}
where $\tilde{Y}_\nu$ is defined in a basis where righthanded neutrinos
are mass eigenstates and $F(x)~\simeq-\frac{3}{2} x$ for small $x$ which
follows from
our assumption that the right handed neutrino masses are hierarchical.
 In order to use this formula, we must
diagonalize
the righthanded neutrino mass matrix and change the $Y_\nu$ to
$\tilde{Y}_\nu$. Since ${\bf M_R}$ is a symmetric complex $2\times 2$
matrix, it can be diagonalized by a transformation matrix
$U(\pi/4)\equiv\frac{1}{\sqrt{2}} \left(\begin{array}{cc}1 & 1\\ -1
& 1\end{array}\right)$ i.e.
$U(\pi/4){\bf M_R}U^T(\pi/4)~=~diag(M_1 , M_2)$ where $M_{1,2}$
are complex numbers. In this basis we have $\tilde{\bf
Y}_\nu~=~U(\pi/4){\bf Y_\nu}$. We can therefore rewrite the formula
for $n_\ell$ as
\begin{eqnarray}
\epsilon_1\propto \sum_j Im [U(\pi/4){\bf Y}_\nu
{\bf Y}^{\dagger}_\nu U^T(\pi/4)]^2_{12}
F(\frac{M_1}{M_2})
\end{eqnarray}
Now note that ${\bf Y}_\nu {\bf Y}^{\dagger}_\nu$ has the form
$\left(\begin{array}{cc}A
& B\\ B & A\end{array}\right)$ which can be diagonalized by the matrix
$U(\pi/4)$. Therefore it follows that $n_\ell =0$.

An interesting feature of this model is that one can determine the
neutrino masses and mixings explicitly in terms of the parameters of the
model. We find a hierarchical mass pattern i.e. $m_1 \ll m_2 \ll m_3$ with
the lightest neutrino being massless i.e.
\begin{eqnarray}
m_1~=~0; m_2~=~\frac{2}{M_+}(h^2_++2h^2_{11}); m_3~=~\frac{2}{M_-}(h^2_-)
\end{eqnarray}
where $M_{\pm}$ are the masses of the two right handed neutrinos with
$M_-\ll M_+$ and $h_{\pm}~=~(h_{22}+h_{23})$.

Even though there is no lepton asymmetry in the model, there are Majorana
CP phases in the light neutrino mixing which we denote by
 $K~=~(e^{i\alpha}, e^{-i\alpha}, 1)$. It is easy to see the origin
of the phases: by appropriate choice of the phases of the fields one can
show that $M_R$ has only one phase and $Y_\nu$ also has only one phase.
After using the seesaw formula, one gets the light neutrino mass matrix
which therefore has only one phase after redefinition of the light
neutrino fields.

\subsection{$\mu-\tau$ SYMMETRY BREAKING WITH TWO RIGHT HANDED
NEUTRINOS} From the above discussion, it is natural to expect the
model to have nonzero lepton asymmetry once $\mu-\tau$ symmetry is
broken, as well as also a nonvanishing $\theta_{13}$. One may then
expect that $\epsilon_1 \propto \theta_{13}$. The details however depend
on
how the symmetry is broken. As an example we note that when the symmetry
is broken only by the
masses of the RH neutrinos i.e. a RH neutrino mass matrix of the
form $M_R~=~diag(M_1,M_2)$ and no off diagonal terms,
 since $Y_\nu Y^{\dagger}_\nu$ is a real matrix, $\epsilon_1
\propto Im[Y_\nu Y^{\dagger}_\nu]_{12}~=~0$ despite the $\mu-\tau$
symmetry breaking. It is easy to check that $\theta_{13}
\simeq c (\theta_A-\frac{\pi}{4})\propto (M_1-M_2)\neq 0$.

One may however break $\mu\leftrightarrow \tau$ symmetry in the
Dirac mass terms for the neutrinos i.e. in $Y_\nu$. This can be
done in many ways e.g. by choosing ${\bf
Y_\nu}~=~\left(\begin{array}{ccc}h_{11}
& h_{22} &  h_{23} \\ h_{12} & h_{23} & h_{22}\end{array}\right)$ or ${\bf
Y_\nu}~=~\left(\begin{array}{ccc}h_{11} & h_{22} &  h_{23} \\ h_{11} &
h_{23} & h_{33}\end{array}\right)$ etc. In all these cases, one gets
$\epsilon_1 \neq 0$ and
also $\theta_{13}\neq 0$ and $\theta_A\neq \pi/4$. 
One lesson one can draw from this observation is that,
 if leptogenesis is the
true mechanism for the origin of matter, then the limit on
$\theta_{13}$ going down by an order of magnitude could teach us
about the nature of right handed neutrino spectrum. For instance, a very
small $\theta_{13}$ (i.e. $\theta_{13}\leq
\frac{\Delta m^2_\odot}{\Delta m^2_A}$) would indicate a nearly
exact $\mu-\tau$ symmetry and therefore sufficient leptogenesis
would then require the existence of three right handed neutrinos or some
complicated way of breaking $\mu-\tau$ symmetry.

 \section{The case of three right handed neutrinos}
In this case, the right handed neutrino mass matrix $M_R$ and the Dirac
Yukawa coupling $Y_\nu$ can be written respectively as:
\begin{eqnarray}
{\bf M_R}~=~\left(\begin{array}{ccc}M_{11} & M_{12} & M_{12}\\ M_{12} &
M_{22} &
M_{23}\\ M_{12} & M_{23} & M_{22}\end{array}\right)\\ \nonumber
{\bf Y_\nu}~=~\left(\begin{array}{ccc}h_{11} & h_{12} & h_{12}\\ h_{21} &
h_{22} & h_{23} \\
h_{21} & h_{23} & h_{22}\end{array}\right)
\end{eqnarray}
where $M_{ij}$ and $h_{ij}$ are all complex\footnote{After this paper was 
posted, it was brought to our attention that leptogenesis for a
$\mu-\tau$ symmetric model with the specific restriction that
$Y_\nu~=~diag(a,b,b)$ was considered in Ref.\cite{gl}. Our
consideration is more general.}. An important property of these two
matrices
is that they can be cast into a block diagonal form by the transformation
matrix $U_{23}(\pi/4)\equiv \left(\begin{array}{cc} 1 & 0\\ 0 &
U(\pi/4)\end{array}\right)$ and then
be subsequently
diagonalized by the most general $2\times 2$ unitary matrix as
follows:
\begin{eqnarray}
V^T(2\times 2)U^T_{23}(\pi/4)M_RU_{23}(\pi/4)V(2\times 2)~=~ M^d_R
\end{eqnarray}
where $V(2\times 2)~=~\left(\begin{array}{cc}V & 0\\ 0 &
1\end{array}\right)$ where $V$ is the most general $2\times 2$
unitary matrix given by $V= e^{i\alpha}P(\beta)R(\theta)P(\gamma)$
with $P(\beta)~=~diag(e^{i\beta}, e^{-i\beta})$;
$R(\theta)~=~\left(\begin{array}{cc}c & s\\ -s &
c\end{array}\right)$; ($c,s$ being cosine and sine of $\theta$
respectively). We will denote $V(2\times 2)$ simply by $V_{L,R}$
depending on whether it acts on left handed or the RH neutrinos.

We now change to the basis where the right handed neutrino mass
matrix is diagonal (Eq.(10)). The Dirac Yukawa coupling in this basis has
the form
\begin{eqnarray}
\tilde{Y}_\nu ~=~V^T(2\times 2)U^T_{23}(\pi/4)Y_\nu\\ \nonumber
\end{eqnarray}
Due to the special form of $Y_\nu$ dictated by $\mu\leftrightarrow\tau$
symmetry, it is easy to see that
\begin{eqnarray}
\tilde{Y}_\nu~=~V^T(2\times 2)Y'_\nu U^T_{23}(\pi/4)
\end{eqnarray}
where $Y'_\nu$ is in block diagonal form.
An important point to realize at this stage is that the the 3$\times $3
matrix problem has reduced to a 2$\times $2 problem. So all the matrices
from now on will be $2\times 2$ and the third neutrino (the heaviest of
the light neutrinos) has completely
``decoupled'' from the considerations below of both seesaw formula
for neutrino masses as well as lepton asymmetry. This is a direct
consequence of $\mu-\tau$ symmetry and of course considerably
simplifies the discussions.

Restricting to the $2\times 2$ case, we can use the seesaw formula to
write down the left handed
neutrino mass matrix as follows in units of $-v^2_{wk}$:
\begin{eqnarray}
{\cal M}_\nu~=~-\tilde{Y}^{T}_\nu M^{d,-1}_R\tilde{Y}_\nu
\end{eqnarray}
Next, we go to a basis where ${\cal M}_\nu$ (the upper $2\times 2$
block of it) is diagonalized by a matrix $V_L$ i.e. $V^T_L{\cal
M}_\nu V_L~=~{\cal M}^d_\nu$. In this basis, the Dirac Yukawa
coupling $\tilde{Y}_\nu$ becomes $V^T_L\tilde{Y}^T_\nu\equiv
Y'_{\nu}{^T}$. Let us write $Y'_{\nu}{^T}$, which is a $2\times 2$
matrix as
\begin{eqnarray}
Y_{\nu}'{^T}~=~ \left(\begin{array}{cc}Z_{11} & Z_{12}\\ Z_{21} &
Z_{22}\end{array}\right)
\end{eqnarray}
The $Z_{ij}$ obey the constraints:
$Z_{12}=-Z_{21}\frac{Z_{22}M_1}{Z_{11}M_2}$ and the neutrino
masses are given by
\begin{eqnarray}
m_1~=~\frac{Z^2_{11}}{M_1}\rho e^{i\eta}\\ \nonumber
m_2~=~\frac{Z^2_{22}}{M_2}\rho e^{i\eta}
\end{eqnarray}
where $\rho e^{i\eta}~=~\left(1+\frac{M_1 Z^2_{21}}{M_2
Z^2_{11}}\right)$.

Let us now calculate the out of equilibrium for the decay of the lightest
right handed neutrino, which we assume to be the lighter of the two mass
eigenstates of the $2\times 2 $ right handed neutrino mass matrix
considered above. It is given by:
\begin{eqnarray}
\Gamma_1 &=& \frac{1}{8\pi} (Y'_\nu Y^{'\dagger}_\nu)_{11} M_1
\nonumber \\ &=&\frac{M_1(|Z_{11}|^2+ |Z_{12}|^2)}{8\pi}\leq 14
\frac{M^2_1}{M_{P\ell}}
\end{eqnarray}
where $M_{P\ell}$ appears in the right hand side from the Hubble expansion
formula $H^2\simeq \sqrt{g_*}T^2/M_{P\ell}$ in a radiation dominated
Universe. Using Eq.($15$), which gives $(|Z_{11}|^2+|Z_{12}|^2)\simeq
\frac{M_1}{v^2_{wk}\rho}\left[|m_1|+|\rho
e^{i\eta}-1||m_2|\right]$, we can rewrite this inequality as a
constraint on the following combination of the masses of the two
lightest neutrino eigenstates:
\begin{eqnarray}
\frac{\left[|m_1|+|\rho e^{i\eta}-1||m_2|\right]}{\rho} \leq 10^{-3}~ eV
\end{eqnarray}
For hierarchical right handed neutrino mass spectrum
(i.e. $M_2\gg M_1$), $\rho\sim 1$ and we get
\begin{eqnarray}
|m_1|+2|m_2||\sin \eta/2|\leq 10^{-3}~ eV
\end{eqnarray}
This puts a limit on the two lightest neutrino masses. For
instance, it implies that the lightest neutrino mass $m_1\leq
10^{-3}$ eV.  The solar neutrino oscillation would require $sin
\eta/2\sim 0.07$ so that $m_2$ will match the central value
required by data.

We now proceed to calculate the primordial lepton asymmetry $\epsilon_1$
in
this model. It turns
out that $\epsilon_1$ is directly proportional to the
solar mass difference square as we show below.
We start with the expression for $\epsilon_1$,
\begin{eqnarray}
\epsilon_1\simeq  \frac{3}{8\pi } \frac{Im(Y'_\nu
Y^{'\dagger}_\nu)^2_{12}}{(Y'_\nu Y^{'\dagger}_\nu)_{11}}
\frac{M_1}{M_2} \nonumber \\ \equiv
\frac{3}{8\pi}\frac{M_1}{M_2}
\frac{Im\left(Z_{11}Z^*_{12}+Z_{21}Z^*_{22}\right)^2}{|Z_{11}|^2+|Z_{12}|^2}.
\end{eqnarray}
Using the constraints on $Z_{ij}$ discussed in Eq. (15) and the relation
just prior to
it, we get, $Im\left(Z_{11}Z^*_{12}+Z_{21}Z^*_{22}\right)^2= |Z_{11}|^4
Im\left(\frac{Z^2_{12}}{Z^2_{11}}\right)+|Z_{22}|^4
\frac{M^2_1}{M^2_2}Im\left(\frac{{Z^2}^*_{12}}{{Z^2}^*_{11}}\right)$. Plugging
this expression into Eq. (19), we can express the primordial lepton
asymmetry $\epsilon_1$ in terms of neutrino masses $m_{1,2}$ and the
parameters $\rho$ and
$\eta$ as follows:
\begin{eqnarray}
\epsilon_1 = \frac{3}{8\pi}\frac{M_1}{v^2_{wk}}\frac{\Delta
m^2_{\odot} sin\eta}{|m_1|+|(\rho e^{i\eta}-1)m_2|} \\ \nonumber   \simeq
 ~10^{-7}\left(\frac{M_1}{10^{10}~GeV}\right)\left(\frac{\Delta
m^2_{\odot}}{8\times 10^{-5}~eV^2}\right)\times \\ \nonumber
\frac{10^{-3}~eV}{\left[|m_1|+|\rho e^{i\eta}-1||m_2|\right]}
(\sin\eta/0.14)
\end{eqnarray}
We see that the origin of matter in this model is predicted primarily in
terms of the solar mass difference square and the unknown phase $\eta$
whose value is already determined by Eq. (18). Thus given a value for
the lightest right handed neutrino mass, the model predicts the value of
primordial lepton asymmetry $\epsilon_1$. In Eq. (20), we have assumed
$M_1\simeq 10^{10}$ GeV. Note that our result is based
on only three assumptions: (i) type I seesaw formula for neutrino
masses and (ii) the existence of $\mu\leftrightarrow \tau$
symmetry and (iii) hierarchy among right handed neutrinos. This is very
different
from generic seesaw models without $\mu\leftrightarrow \tau$ where
the dominant contribution to $\epsilon_1$ comes from the atmospheric
neutrino mass difference square and depends on unknown parameters related
to the Dirac neutrino Yukawa coupling\cite{buch}. It is also interesting
that origin of matter is tied to
the existence of solar neutrino oscillation and it is the LMA
solution to the solar neutrino problem that reproduces the correct
order of magnitude for the lepton asymmetry which after taking
into the dilution factor\cite{KTPbuch} and sphaleron effects, can
give rise to the magnitude for the observed baryon to photon
ratio. The value of $10^{10}$ GeV for the mass of the lightest
right handed neutrino is chosen to show that the model when
embedded into an extension of MSSM can avoid the reheat
temperature constraint coming from gravitino production.
 Finally it is important to stress that this result is valid for both
normal and inverted mass hierarchy among light neutrinos.

In conclusion, we have discussed the consequences of the hypothesis that
the large atmospheric neutrino mixing angle arises from an intrinsic
$\mu-\tau$ symmetry for leptons for origin of matter via leptogenesis.
 We point out that if there are
two right handed neutrinos obeying $\mu-\tau$ interchange symmetry, then
lepton asymmetry vanishes whereas for three right handed
neutrinos, it is given directly the solar mass difference square provided
one assumes type I seesaw formula for neutrino masses. We also obtain an
upper limit on the lightest neutrino mass of a milli-eV under these
assumptions.

 This work is supported by the National Science Foundation grant
no. Phy-0354401.

\end{document}